\documentclass[prl,twocolumn,amsmath,amssymb,showpacs]{revtex4}

\usepackage{graphicx}
\usepackage{dcolumn}
\usepackage{bm}
\usepackage{mathrsfs}
\usepackage{appendix}
\usepackage{bbm}
\usepackage{color}

\def\be{\begin{equation}}
\def\ee{\end{equation}}
\def\bd{\begin{displaymath}}
\def\ed{\end{displaymath}}
\def\-{\phantom{-}}
\def \vR {{\bf R}}
\def \vk {{\bf k}}
\def \vq {{\bf q}}
\def \vx {{\bf x}}
\def \vy {{\bf y}}
\def \vz {{\bf z}}

\begin{document}

\title{Spin-Wave Instabilities and Non-Collinear Magnetic Phases of a Geometrically-Frustrated Triangular-Lattice Antiferromagnet}
\author{J.T. Haraldsen,$^1$ M. Swanson,$^{1,2}$ G. Alvarez,$^3$ and R. S. Fishman$^1$}
\affiliation{$^1$Materials Science and Technology Division, Oak Ridge National Laboratory, \mbox{Oak Ridge, TN 37831}}
\affiliation{$^2$Department of Physics, North Dakota State University, \mbox{Fargo, ND 58105}}
\affiliation{$^3$Computer Science \& Mathematics 
Division and Center for Nanophase Materials Sciences, Oak Ridge National Laboratory, \mbox{Oak Ridge, TN 37831}}

\begin{abstract}

This paper examines the relation between the spin-wave instabilities of 
collinear magnetic phases and the resulting non-collinear phases for a geometrically-frustrated 
triangular-lattice antiferromagnet in the high spin limit.  Using a combination of phenomenological and 
Monte-Carlo techniques, we demonstrate that the instability wave-vector with the strongest intensity in the 
collinear phase determines the wave-vector of a cycloid or the dominant elastic peak of a more 
complex non-collinear phase.  Our results are related to the observed multi-ferroic phase of
Al-doped CuFeO$_2$.
\end{abstract}

\pacs{75.30.Ds, 75.50.Ee, 61.05.fg}

\maketitle

It is well-known that the transition between different magnetic ground states may be signaled by the softening of a 
spin-wave (SW) mode.  In the simplest case of a conventional square-lattice antiferromagnet, the softening of a SW mode at wave-vectors 
$(\pi ,0)$ and $(0,\pi )$ signals the spin flop and canting of the magnetic moments at a critical field.  
In the manganites \cite{fang:00}, the SW instabilities of the ferromagnetic state have been used to construct the phase
diagram for the antiferromagnetic (AF) phases that appear with Sr doping.  The softening of a SW excitation at $(\pi ,0)$ 
signals the instability of the N\'eel state and the canting of the spins in a spin-1/2 union-jack lattice \cite{zheng:07}.
But the relation between the SW instabilities of a collinear phase and a resulting non-collinear phase is 
less clear when multiple SW instabilities occur simultaneously or when the non-collinear phase 
has a complex magnetic structure with several elastic peaks.  This paper explores the relation between the SW instabilities of 
the collinear 4 and 8-sublattice (SL) phases of a geometrically-frustrated triangular-lattice antiferromagnet (TLA) and 
the non-collinear phases that appear with decreasing anisotropy $D$.  When multiple 
SW instabilities of the collinear phase occur at once, the instability wave-vector with the largest 
intensity determines the dominant ordering wave-vector of the resulting non-collinear phase.  
One of the predicted non-collinear phases may be related to the multi-ferroic phase that appears in CuFeO$_2$ with Al doping \cite{ter:04}.

\begin{figure}
\includegraphics[width=3.25in]{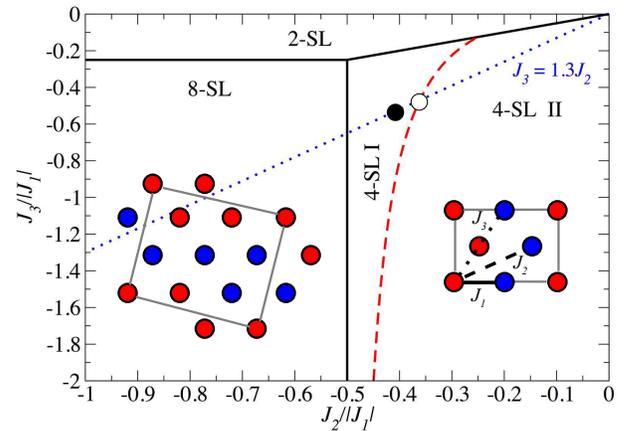}
\caption{(Color online) A portion of the TLA phase diagram for large $D$ and $J_1 <0$.  
The dashed (red) curve divides the 4-SL phase into 
4-SL I and 4-SL II regions.  For the 4 and 8-SL phases, red circles denote 
up spins and blue circles denote down spins;  the solid (gray) line defines the unit cell.  
The dashed (blue) line obeys the relation $J_3 = 1.3J_2$, the black circle is the estimated location of the exchange parameters for 
CuFeO$_2$, and the white circle lies on the boundary between the 4-SL I and 4-SL II regions. }
\label{phases}
\end{figure}

Frustrated TLAs with AF nearest-neighbor exchange $J_1 < 0$ exhibit a remarkable number of competing ground states \cite{die:04}.  
With interactions $J_i$ up to third nearest neighbors (denoted in Fig. \ref{phases}) and assuming Ising 
spins along the $\vz$ direction, Takagi and
Mekata \cite{tak:95} obtained a phase diagram with ferromagnetic (FM), 2-SL, 3-SL, 4-SL, and 8-SL phases.
A portion of that phase diagram is sketched in Fig. \ref{phases}.  For the geometrically-frustrated TLA 
CuFeO$_2$ in fields below 7 T, the ground state is the 4-SL phase \cite{mit:91, mek:93} and the black dot in Fig. \ref{phases}
denotes the estimated ratio of exchange parameters $J_2/\vert J_1\vert \approx -0.44$ and $J_3/\vert J_1\vert \approx -0.57$ \cite{ye:07, fis:08b}.

As demonstrated by the small SW gap of about 0.9 meV \cite{ye:07, ter:04} on either side of 
the ordering wave-vector $\vq =\pi \vx$, the spin fluctuations of CuFeO$_2$ are
much softer than would be expected for Ising spins.  With Heisenberg spins, the collinear magnetic phases of a TLA 
become locally unstable below a critical anisotropy $D_c$ that depends on the exchange parameters $J_i$.  The 
observed softening of the SW modes in CuFe$_{1-x}$Al$_x$O$_2$ with Al doping \cite{ter:04} can be reproduced 
by lowering $D$ towards $D_c$ \cite{fis:08a} in the 4-SL I region of Fig. \ref{phases}.  For Al concentrations above about 
$x_c \approx 0.016$, the magnetic ground state of CuFe$_{1-x}$Al$_x$O$_2$ becomes 
non-collinear and displays multi-ferroic properties \cite{kim:06, kan:07, sek:07}.

We have determined the ground-state magnetic phases of the TLA using a combination of Monte-Carlo (MC)
simulations and phenomenological techniques.  The TLA Hamiltonian is
\begin{equation}
H=-\frac{1}{2} \sum_{i\neq j} J_{ij}\mathbf{S}_i \cdot \mathbf{S}_j - D\sum_i S^2_{iz} ,
\label{Ham}
\end{equation}
where $J_{ij}$ includes first, second, and third-neighbor interactions (shown in Fig. \ref{phases}).   
The nearest-neighbor distance has been set to 1.
The SW frequencies of the collinear phases are obtained by performing a 
Holstein-Primakoff $1/S$ expansion about the classical limit.  Above the critical
anisotropy $D_c$, a collinear phase is locally stable if the SW frequencies $\omega (\vk )$
are positive and real for every momentum $\vk $.

MC simulations were used to find the non-collinear magnetic
phases of the TLA.  The simulations were started at a high-enough temperature 
to rule out metastable states.  To mimic the process of thermal annealing, the system was
slowly cooled to a final temperature (in units of $\vert J_1\vert S^2$) ranging from $4\times 10^{-3}$ to $1\times 10^{-4}$.
Lowering the final temperature further did not significantly change the resulting 
non-collinear phase.  Using lattices of varying sizes with periodic boundary conditions, 
we found that there was no substantial change for lattices greater than $16\times16$.

\begin{figure}
\includegraphics[width=3.25in]{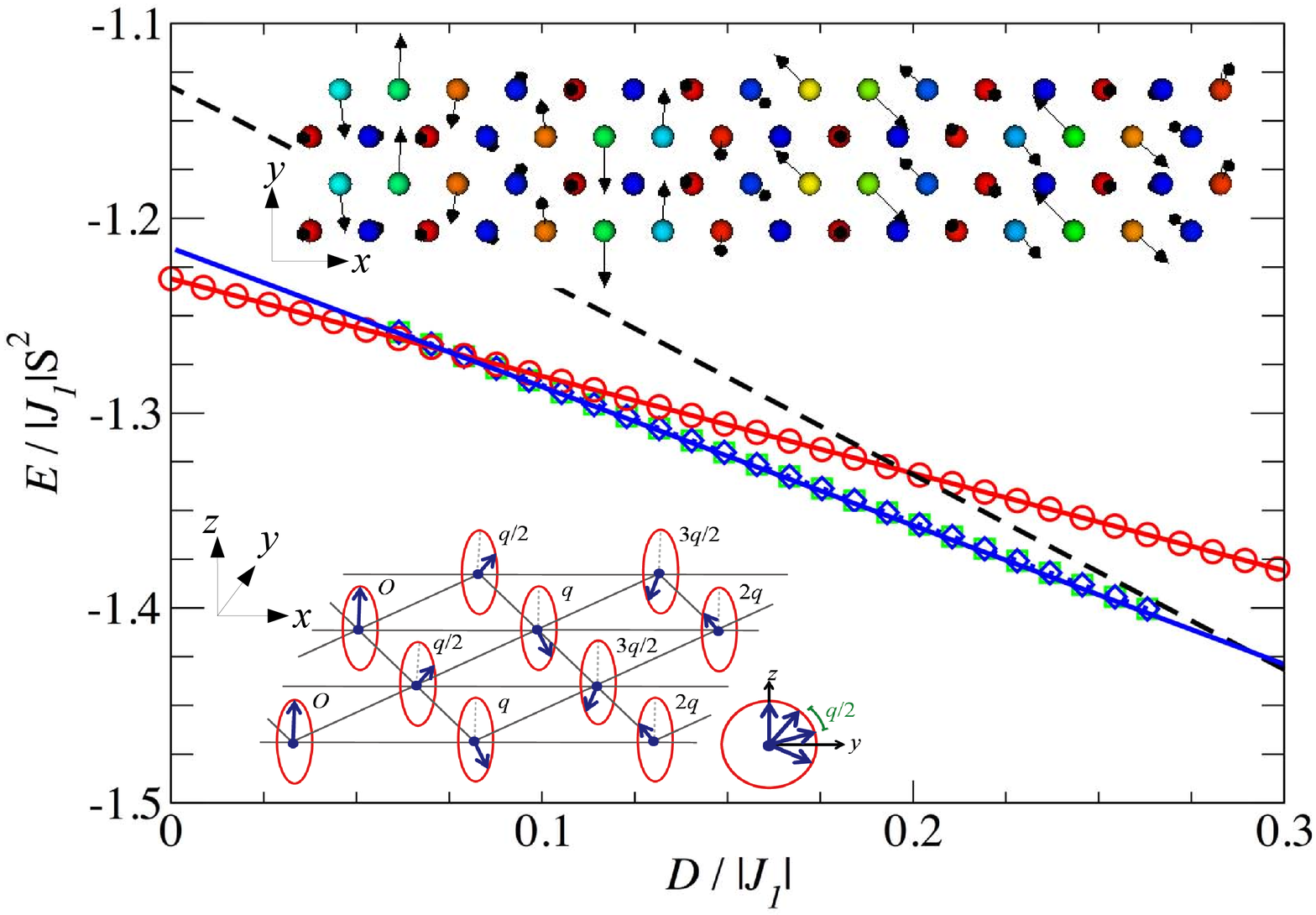}
\caption{(Color online) Energy as a function of $D/\vert J_1\vert $ for the 4-SL (black dashed), CNC (blue diamonds), and cycloid I 
(red circles) phases with $J_3/J_2=1.3$ and $J_1/J_2=2.28$.  The bottom diagram shows the 
the cycloid for arbitrary $q$. The top diagram shows the CNC phase with up spins in red and down spins in blue.}
\label{EvsD}
\end{figure}

In Fig. \ref{phases}, the 4-SL phase is separated into regions I and II by the curve $J_1/J_2-2= J_2/J_3$.  
In region 4-SL II, the instability wave-vectors are given
by $\vk = (\pi \pm \pi /3)\vx $, independent of the exchange parameters;  in region 4-SL I, the instability
wave-vectors depend on the exchange parameters \cite{swan:un}.  With the exchange parameters corresponding to 
the black circle in region 4-SL I, we determined the stable magnetic phases as a function of $D$.  As shown in Fig. \ref{EvsD},
the 4-SL phase is stable down to $D/\vert J_1\vert \approx 0.27$, below which MC simulations obtain
the complex non-collinear (CNC) phase shown on the top of Fig. \ref{EvsD}.  Since the same up or down spin
frequently occurs at sites $\vR =m\vx + n\sqrt{3}\vy $ and $\vR' = \vR + \vx /2 + \sqrt{3} \vy /2$ or 
$\vR^{\prime \prime} = \vR - \vx /2 + \sqrt{3} \vy /2$, the CNC phase retains some of the FM correlations present in the 
4-SL phase.  Translationally invariant in the $\vy $ direction, the CNC phase has a period in the $\vx $ direction between 2 and 3 lattice constants.  
Becauase the MC simulations were performed on a finite lattice, the energy of the CNC phase is overestimated and we 
cannot say whether this phase is commensurate or incommensurate.  

Below a second threshold value of $D/\vert J_1\vert \approx 0.08$, a cycloid like the one
sketched in the bottom panel of Fig. \ref{EvsD} \cite{cycl} has a lower energy than the CNC phase.  As discussed below, 
the wave-vector of the cycloid is independent of $D$.  If the CNC phase were neglected, then the cycloid would 
achieve a lower energy than the 4-SL phase below $D/\vert J_1\vert \approx 0.2$, still above the critical 
value $D_c/\vert J_1\vert \approx 0.15$ for the local stability of the 4-SL phase.  

\begin{figure}
\includegraphics[width=3.5in]{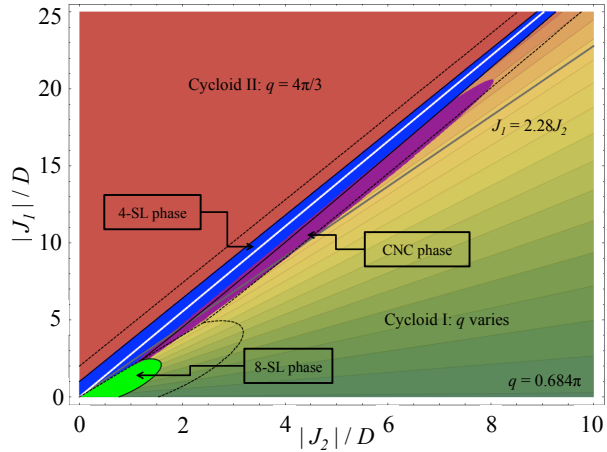}
\caption{(Color online) Phase diagram for the TLA as a function of $|J_1|$/$D$ and $|J_2|$/$D$ with $J_3$ = 1.3$J_2$ 
containing five regions:  4-SL (blue), 8-SL (green), CNC (violet), cycloid I (variable green-orange), and cycloid II (maroon).
The dashed (white) line separates regions 4-SL I and 4-SL II. 
The dotted (black) curves denote the metastable boundaries for the 4-SL and 8-SL regions. Cycloid I has 
wave-vectors $q$ that range from 0.684$\pi$ to 0.923$\pi$ in intervals of 0.016$\pi$.}
\label{PD-phases}
\end{figure}

To gain a better understanding of the phases stabilized within the TLA, we
have evaluated the magnetic phases along the line with $J_3/J_2=1.3$ drawn through the black dot in Fig. \ref{phases}.
Five stable phases are presented as a function of $\vert J_1\vert /D$ and $\vert J_2\vert /D$ in Fig. \ref{PD-phases}.  
The 4-SL phase is stable along a strip through the diagonal
of this plot.  Although not indicated by this figure, the 4-SL region disappears above $\vert J_1\vert /D\approx 40$. 
Close to the origin or for large $D$, a collinear 8-SL region is indicated in Fig. \ref{phases}. 
Two cycloids are also obtained:  in the upper left, cycloid II with wave-vector $4\pi \vx /3$;  in the lower right, 
cycloid I with the variable wave-vector indicated in the figure \cite{cycl}.  Finally, a CNC phase appears just below the 
4-SL phase and disappears above $\vert J_1\vert /D \approx 20$.   
Regions of local stability for the collinear phases are indicated in Fig. \ref{PD-phases} by the dashed black lines \cite{swan:un}.  
The results in Fig. \ref{EvsD} can be obtained from Fig. \ref{PD-phases} by drawing a line from the 
origin with slope 2.28 (gray) (so that $J_2/\vert J_1\vert =-0.44$), which passes from the 
4-SL phase through the CNC phase into cycloid I.  

The classical energies of each of these phases can be written as
$E/S^2 = A_1J_1 + A_2J_2 + A_3J_3 - A_DD$.  The coefficients for each phase are given in 
Table \ref{coefficients}.  Only a non-collinear phase with $0.5 < A_D < 1$ can intercede 
between a collinear phase with $A_D=1$ and a cycloid with $A_D=0.5$.  For the CNC phase with $A_D\approx 0.71$, the error bars
indicate the range of parameters obtained from MC simulations near $\vert J_1\vert /D = 5.7$ and $\vert J_2\vert /D = 2.5$.  
This phase is characterized by rather weak next-neighbor correlations with small $\vert A_2\vert $.
The CNC phase space in Fig. \ref{PD-phases} is obtained by using the results of Table \ref{coefficients}
to evaluate the energies of the MC spin configurations as functions of $\vert J_1\vert /D$ and $\vert J_2\vert /D$.  Hence,
the CNC region may be {\it underestimated}.  

The ordering wave-vector $\vq =q\vx $ of cycloid I is evaluated by minimizing $E$ with respect to $q$.
So $q$ depends only on the ratios $J_2/J_1$ and $J_3/J_2$, as indicated by the diagonal lines in 
Fig. \ref{PD-phases}.   Cycloid II with $q=4\pi /3$ corresponds to the 
120$^o$ N\'eel state found in a classical TLA with $D=0$ and nearest and next-nearest neighbor 
interactions \cite{jol:90} when $\vert J_2/J_1\vert < 1/8$.  A slightly distorted N\'eel state \cite{cycl}
is stable for nonzero $D$ over a range of exchange parameters with $\vert  J_3/J_2\vert > 1/2$, so that the 
diagonal line in Fig. \ref{phases} passes through the 4-SL and 8-SL phases.  MC simulations were used to confirm
the stability of cycloids I and II in Fig. \ref{PD-phases}.  

\begin{table*}
\caption{\textbf{Energy Coefficients for Collinear, Cycloid, and CNC Phases}}
\begin{ruledtabular}
\begin{tabular}{lcccc}
 \textbf{Phase} & $A_1$  & $A_2$ & $A_3$ & $A_D$  \\
\hline 
4-SL & 1 & -1 & 1 & 1 \\
8-SL & 0 & 1 & 1 & 1 \\
Cycloid I& -$\big(\cos(q)+2\cos(q/2)\big)$ & -$\big(1+2\cos(3q/2)\big)$ & -$\big(\cos(2q) + 2\cos(q)\big)$ & $1/2$ \\
Cycloid II& 3/2 & -3 & 3/2 & 1/2 \\
CNC & 0.595$\pm$0.001 & -0.097$\pm$0.001 & 1.161$\pm$0.001 & 0.712$\pm$0.001 \\
\end{tabular}
\end{ruledtabular}
\label{coefficients}
\end{table*}

\begin{figure}[b]
\includegraphics[width=3.5in]{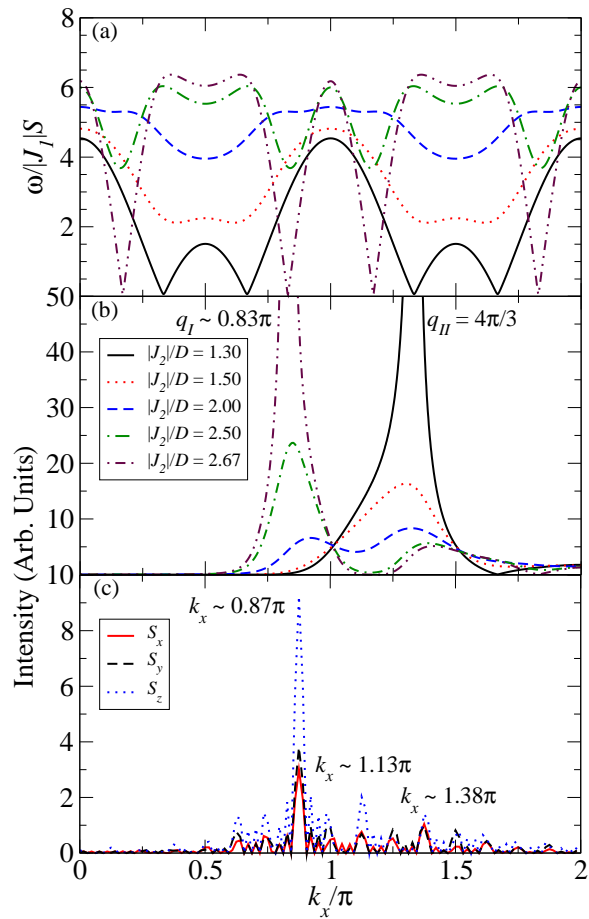}
\caption{(Color online)(a-b) SW frequency and intensity versus wave-vector $k_x$ for the 4-SL phase with $|J_1|/D$ = 5.5, $J_3 = 1.3J_2$, where
$\vert J_2\vert /D$ varies from 1.30 to 2.67.  (c) Fourier transform for the $S_x$, $S_y$, and 
$S_z$ components of the CNC phase with the same exchange parameters as above and $|J_2|/D =2.5$. }
\label{CNC-Fourier}
\label{4-SLwaves}
\end{figure}

With decreasing $D$ or moving away from the origin of Fig. \ref{PD-phases} along a diagonal, the 4-SL phase becomes
unstable either to cycloid II or to the CNC phase.  The white line bisecting region the 4-SL strip in Fig. \ref{PD-phases} corresponds
to the white point in Fig. \ref{phases} at the border between the 4-SL I and 4-SL II regions with $J_2/\vert J_1\vert = -0.36$.
In region 4-SL II or above the white diagonal line, the 
4-SL phase has instabilities at the wave-vectors $(\pi \pm \pi /3)\vx $.  The SW intensity
at the larger of these two wave-vectors always dominates and the 4-SL phase evolves into 
cycloid II with wavector $4\pi \vx /3$.

In region 4-SL I, the 4-SL phase has three unique SW instabilities:  one at wave-vector $\vk_1$ along the
$\vx $ axis, another at $\vk_2$ rotated by $\pi /3$, and a third at $\vk_3$ rotated by $-\pi /3$.  All three have the
same magnitude with $\pi /2 < k_i < \pi $.  Other SW instabilities in the 4-SL I region can be related 
by a symmetry operation to one of these three.  We find that the instability at $\vk_1$ always has a larger intensity
than the ``twins'' at $\vk_2 $ or $\vk_3$ or than any of the other wave-vectors related by symmetry.
Correspondingly, cycloid I along any diagonal in Fig. \ref{PD-phases} has the same wave-vector $\vq $ as the 
dominant instability of the 4-SL phase.   

Similar conclusions are reached for the 8-SL phase, which switches to cyloid I
along any diagonal in Fig. \ref{PD-phases}.  Although the SW instability of the 8-SL phase occurs simultaneously at
several wave-vectors, the dominant wave-vector instability of the 8-SL phase
coincides with the wave-vector $\vq $ of cycloid I along any diagonal in Fig. \ref{PD-phases}.

However, the CNC phase that intercedes between the 4-SL and cycloid I phases 
is characterized by several elastic peaks shown in Fig. \ref{CNC-Fourier}(c).  Within the precision of our MC simulations,
the dominant wave-vector $k_x \approx 0.87 \pi $ of the CNC phase
coincides with the dominant instability wave-vector of the 4-SL phase that preceeds it. 

To demonstrate how the magnetic ground state evolves from cycloid II into the 4-SL phase and then into cycloid I,
we plot in Fig. \ref{4-SLwaves}(a) and (b) the SW frequencies and intensities versus wave-vector for the 4-SL phase with 
$J_1/D = -5.5$, $J_3/J_2 = 1.3$, and $\vert J_2\vert /D$ 
varying from 1.30 to 2.67.  As $|J_2|/D$ approaches the lower limit for the stability of the 4-SL phase,
the SW intensity dominates at the cycloid II wave-vector $\vq = 4\pi \vx /3$.   At the upper limit, 
the SW intensity dominates at the cylcoid I wave-vector $\vq \approx 0.83 \pi \vx $. Hence,
the wave-vector of the SW instabilities for the collinear phases correspond to 
the ordering wave-vectors of the non-collinear phases.   

Bear in mind that the transition between magnetic ground states is not always signaled
by the softening of a SW mode.  In fact, the transition between the 4-SL and 8-SL phases in Fig. \ref{phases}
occurs even at $D=\infty $, when the spins are Ising variables and the SW gaps for both phases diverge.

The CNC phase may be related to the multi-ferroic phase observed in 
Al-doped CuFeO$_2$ \cite{ter:04}, which was recently 
investigated by Nakajima {\it et al.} \cite{nak:07}.  Based on neutron-scattering measurements, those
authors concluded that the ground state is a modified cycloid with the same spin on 
sites $\vR $ and $\vR' $ (see above).  
This phase has peaks at wave-vectors on either side of $\pi\vx $, in agreement with the neutron
measurements.  However, a modified cycloid cannot be stabilized
by a Hamiltonian with the form of Eq.(\ref{Ham}), regardless of the exchange and anisotropy parameters.   
With an additional phase slip $\delta $ for the spins at sites $\vR'$, a pure cycloid with 
$\delta =0$ and a single elastic peak always has lower energy than the phase proposed in Ref.\cite{nak:07}
with $\delta = -q/2$.  This conclusion has been verified by MC simulations.

Like the non-collinear phase proposed earlier \cite{nak:07}, the CNC phase also contains FM correlations
between sites $\vR $ and $\vR'$ or $\vR''$.  So the CNC phase also has elastic peaks on either side of 
$\pi \vx$ at $k_x \approx 0.87 \pi $ and $1.13 \pi $, as shown in Fig. \ref{CNC-Fourier}(c). 
Because the FM correlations are not perfect and vary along the $\vx $ direction,
the CNC phase contains several other elastic peaks that may allow it to 
be experimentally distinguished from the phase proposed in Ref.\cite{nak:07}.  

To summarize, we have shown that the dominant wave-vector of a non-collinear phase in a frustrated
TLA corresponds to the dominant instability wave-vector of a
collinear phase as the anisotropy is lowered and spin fluctuations become softer.
The CNC phase sketched in Fig. \ref{EvsD} is a more reasonable candidate for the
multi-ferroic phase observed in Al-doped CuFeO$_2$ than the one previously proposed.

This research was sponsored by the Laboratory Directed
Research and Development Program of Oak Ridge National Laboratory,
managed by UT-Battelle, LLC for the U. S. Department of Energy
under Contract No. DE-AC05-00OR22725 and by the Division of Materials Science
and Engineering and the Division of Scientific User Facilities of the U.S. DOE.


\begin{thebibliography}{PD}
\bibitem{cycl}  Due to the anisotropy, the cycloids will be slightly distorted by the tilting of the spins 
towards the $\pm \vz$ directions with an energy gain of order $D^2S^2/z\vert J_1 \vert $ ($z=6$ is the 
number of nearest neighbors).  Within the precision of our MC simulations, this distortion is undetectable and 
a pure cycloid is obtained.
\end{thebibliography}
\end{document}